\pdfoutput=1

\newif\ifDraft

\documentclass[graybox]{svmult}

\usepackage{mathptmx}       
\usepackage{helvet}         
\usepackage{courier}        
\usepackage{type1cm}        
\usepackage{makeidx}         
\ifDraft
\usepackage[draft]{graphicx}
\else
\usepackage[final]{graphicx}
\fi
\usepackage{multicol}        
\usepackage[bottom]{footmisc}
\usepackage{url}
\usepackage{cite}
\usepackage{amssymb}
\usepackage{amsmath}

\makeindex             

\graphicspath{
{figures/}
}

\usepackage{comment}

\PassOptionsToPackage{pdftex,final}{graphicx}
\PassOptionsToPackage{dvipsnames}{xcolor}
\usepackage{tikz}
\usetikzlibrary{snakes}

\newlength{\mylength}

\title*{From individual to population:\\Challenges in Medical Visualization}
\titlerunning{Challenges in Medical Visualization}

\author{C.P. Botha, B. Preim, A. Kaufman, S. Takahashi, A. Ynnerman}

\authorrunning{Botha, Preim, Kaufman, Takahashi, Ynnerman}

\institute{
C.P. Botha \at Computer Graphics and Visualization, Delft University
of Technology, also Division of Image Processing, Leiden University Medical
Center, The Netherlands, \email{c.p.botha@tudelft.nl} \and
B. Preim \at Department of Simulation and Graphics, University of Magdeburg,
Germany, \email{preim@isg.cs.uni-magdeburg.de} \and 
A. Kaufman \at Department of Computer Science, Stony Brook University, New
York, USA, \email{ari@cs.stonybrook.edu} \and 
S. Takahashi \at Graduate School of Frontier Sciences, 
The University of Tokyo, Japan, \email{takahashi@acm.org} \and
A. Ynnerman \at Norrk\"oping Visualization and Interaction Studio, Link\"oping
University, Sweden, \email{anders.ynnerman@itn.liu.se} }

\begin{document}

\maketitle

\textbf{Note:} Also to appear in the Dagstuhl 2012 SciVis book by
Springer. Please cite this paper with its arXiv citation information.
 
%

\section{Introduction}

Since the advent of magnetic resonance imaging (MRI) and computed
tomography (CT) scanners around the early seventies, and the consequent
ubiquitousness of medical volume data, medical visualization has
undergone significant development and is now a primary branch of
Visualization. It finds application in diagnosis, for example virtual
colonoscopy, in treatment, for example surgical planning and guidance,
and in medical research, for example visualization of diffusion tensor
imaging data. Although the field of medical visualization only
established itself with this name in the late
eighties~\cite{preim_visualization_2007}, we shall see in the next
section that already in the seventies there were published examples of
computer-generated images, based on medical data and used for medical
applications.

During the past decades, medical image acquisition technology has undergone
continuous and rapid development. It is now possible to acquire
much more complex data than ever before. For example, in High Angular
Resolution Diffusion Imaging (HARDI), forty or more diffusion-weighted
volumes are acquired in order to calculate and visualize water diffusion
and, indirectly, structural neural connections in the
brain~\cite{tuch_high_2002}.
In fMRI-based full brain connectivity, time-based correlation of neural
activity is indirectly measured between all pairs of voxels in the brain, thus
giving insight into the functional neural
network~\cite{greicius_functional_2003}. Moreover, the questions that users
attempt to answer using medical visualization have also become significantly
more complex.

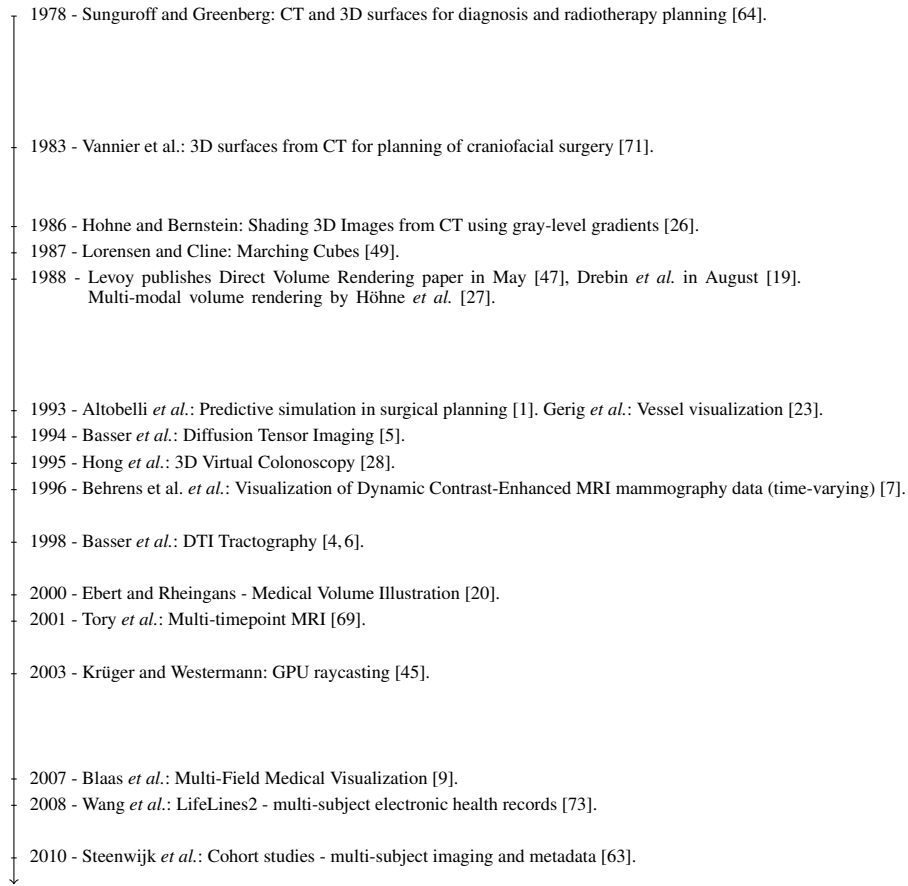
\begin{figure}
\begin{tikzpicture}[scale=0.35, snake=zigzag, line before snake = 5mm, line
after snake = 5mm]
\def\sa{1978}
\def\ea{2011}
\def\ad{\ea-\sa}
\newcommand{\tlfont}{\scriptsize}

\draw[->] (0,0) -- (0,\sa-\ea);

\draw (0,0) node[right=3pt] { \tlfont 1978 -
Sunguroff and Greenberg: CT and 3D surfaces for diagnosis and
radiotherapy planning~\cite{sunguroff_computer_1978-1}. }; \draw (-3pt,0) --
(3pt,0);

\def\y{\sa-1983}
\draw (0,\y) node[right=3pt] { \tlfont 1983 - Vannier et al.: 3D
surfaces from CT for planning of craniofacial
surgery~\cite{vannier_three_1983}. }; \draw (-3pt,\y) -- (3pt,\y);

\def\y{\sa-1986}
\draw (0,\y) node[right=3pt] { \tlfont 1986 - Hohne and Bernstein:
Shading 3D Images from CT using
gray-level gradients~\cite{hohne_shading_1986}. }; \draw (-3pt,\y) --
(3pt,\y);

\def\y{\sa-1987}
\draw (0,\y) node[right=3pt] { \tlfont 1987 - Lorensen and Cline: Marching
Cubes~\cite{lorensen_marching_1987}. }; \draw (-3pt,\y) -- (3pt,\y);

\def\y{\sa-1988}
\draw (0,\y) node[right=3pt, text width=0.98\textwidth] { \tlfont 1988 - Levoy
publishes Direct Volume Rendering paper in May~\cite{levoy_display_1988}, Drebin {\em et al.}\ in August~\cite{drebin_volume_1988}.}; \draw (-3pt,\y) -- (3pt,\y);

\def\y{\sa-1988-0.75}
\draw (0,\y) node[right=3pt, text width=0.98\textwidth] { \tlfont
\newlength{\eewidth}\settowidth{\eewidth}{{1988 -}}\hspace{\eewidth} 
Multi-modal volume rendering by H\"ohne {\em et
al.}~\cite{hohne_display_1988}. };

\def\y{\sa-1993}
\draw (0,\y) node[right=3pt] { \tlfont 1993 - Altobelli {\em et al.}:
Predictive simulation in surgical
planning~\cite{altobelli_computer-assisted_1993}. Gerig {\em et al.}: Vessel
visualization~\cite{barrett_symbolic_1993}.}; \draw (-3pt,\y) -- (3pt,\y);

\def\y{\sa-1994}
\draw (0,\y) node[right=3pt] { \tlfont 1994 - Basser {\em et al.}:
Diffusion Tensor Imaging~\cite{basser_mr_1994}. };
\draw (-3pt,\y) -- (3pt,\y);

\def\y{\sa-1995}
\draw (0,\y) node[right=3pt] { \tlfont 1995 - Hong {\em et al.}: 3D Virtual
Colonoscopy~\cite{lichan_hong_3d_1995}. };
\draw (-3pt,\y) -- (3pt,\y);

\def\y{\sa-1996}
\draw (0,\y) node[right=3pt] { \tlfont 1996 - Behrens et al. {\em et al.}:
Visualization of Dynamic Contrast-Enhanced MRI
mammography data (time-varying)~\cite{behrens_computer-assisted_1996}. }; \draw (-3pt,\y)
-- (3pt,\y);

\def\y{\sa-1998}
\draw (0,\y) node[right=3pt] { \tlfont 1998 - Basser {\em et al.}:
DTI Tractography~\cite{basser_fiber-tractography_1998, basser_vivo_2000}. };
\draw (-3pt,\y) -- (3pt,\y);

\def\y{\sa-2000}
\draw (0,\y) node[right=3pt] { \tlfont 2000 - Ebert and Rheingans - Medical Volume
Illustration~\cite{ebert_volume_2000}. }; \draw (-3pt,\y) -- (3pt,\y);

\def\y{\sa-2001}
\draw (0,\y) node[right=3pt] { \tlfont 2001 - Tory {\em et al.}:
Multi-timepoint MRI~\cite{tory_4d_2001}. }; \draw (-3pt,\y) -- (3pt,\y);

\def\y{\sa-2003}
\draw (0,\y) node[right=3pt] { \tlfont 2003 - Kr\"uger and Westermann:
GPU raycasting~\cite{kruger_acceleration_2003}. }; \draw (-3pt,\y) -- (3pt,\y);

\def\y{\sa-2007}
\draw (0,\y) node[right=3pt] { \tlfont 2007 - Blaas {\em et al.}:
Multi-Field Medical Visualization~\cite{blaas_interactive_2007}. }; \draw (-3pt,\y) --
(3pt,\y);

\def\y{\sa-2008}
\draw (0,\y) node[right=3pt] { \tlfont 2008 - Wang {\em et al.}:
LifeLines2 - multi-subject electronic health
records~\cite{wang_aligning_2008}. }; \draw (-3pt,\y) -- (3pt,\y);

\def\y{\sa-2010}
\draw (0,\y) node[right=3pt] { \tlfont 2010 - Steenwijk {\em et al.}:
Cohort studies - multi-subject imaging and metadata~\cite{steenwijk_integrated_2010}.
}; \draw (-3pt,\y) -- (3pt,\y);

\end{tikzpicture}
\caption{Timeline with a subset of medical visualization papers showing the
progression from scalar volume datasets through time-dependent data to
multi-field and finally multi-subject datasets. This timeline is by no means
complete, instead attempting to show a representative sample of papers that
represent various trends in the development of the field.}
\label{fig:medvis_timeline}
\end{figure}

In this paper, we first give a high-level overview of medical
visualization development over the past 30 years, focusing on key
developments and the {\em trends} that they represent. During this
discussion, we will refer to a number of key papers that we have also
arranged on the medical visualization research timeline shown in
figure~\ref{fig:medvis_timeline}. Based on the overview and our
observations of the field, we then identify and discuss the medical
visualization research challenges that we foresee for the coming decade.


\section{Thirty-year Overview of Medical Visualization}
\label{sec:overview}

Already in 1978, Sunguroff and Greenberg published their work on the
visualization of 3D surfaces from CT data for diagnosis, as well as a
visual radio-therapy planning system, also based on CT
data~\cite{sunguroff_computer_1978-1}. Five years later, Vannier {\em et
al.} published their results developing a system for the computer-based
pre-operative planning of craniofacial
surgery~\cite{vannier_three_1983}. By this time, they had already used
and evaluated their surgical planning system in treating 200 patients.
The system was based on the extraction and visualization of 3D hard and
soft tissue surfaces from CT data. Through the integration of an
industrial CAD application, it was also possible to perform detailed 3D
measurements on the extracted surfaces.

\subsection{Practical and Multi-modal Volume Visualization}
In 1986, Hohne and Bernstein proposed using the gray-level gradient to perform
shading of surfaces rendered from 3D CT data~\cite{hohne_shading_1986}. In
1987, Lorensen and Cline published the now famous Marching Cubes
isosurface extraction algorithm, which enabled the fast and practical extraction of 3D
isosurfaces from real-world medical data. In the year thereafter, Levoy
introduced the idea of volume raycasting in May~\cite{levoy_display_1988}, and
Drebin {\em et al.}\ in August~\cite{drebin_volume_1988}. Although medical
volume visualization was possible before these publications, as witnessed by a
number of publications, previous techniques were either not as fast or yielded
less convincing results. With the introduction of Marching Cubes and volume
raycasting, volume visualization became a core business of visualization and
medical visualization for the years to come.

Up to this point, research had focused on uni-modality data, primarily CT.
However, already in 1988 the first multi-modal volume rendering paper was
published by H\"ohne {\em et al.}, in which they demonstrated the registration
and combined visualization of CT and MRI. A great deal of work has been
done since then on the theory and applications of multi-modal volume
visualization. The first notable example is the work of Cai and Sakas in
1999 where they classified voxel-voxel multi-modal volume rendering
techniques according to the volume rendering pipeline stage where they
take place~\cite{cai_data_1999}. The three classes are image level,
where two volume renderings are combined pixel-by-pixel,
accumulation level, where looked up samples along the ray are combined,
and illumination model level, where the illumination model is adapted
to process two volume samples directly. The second example we mention
is a convincing application of multi-modal volume rendering for the
planning of neurosurgical interventions, where MRI, CT, fMRI, PET and
DSA data are all combined in an interactive but high quality
visualization for the the planning of
brain tumor resection~\cite{beyer_high-quality_2007}.



\subsection{Therapy Planning, Predictive Simulation, and Diagnosis}
\label{sec:overview_prediction}
Therapy planning was one of the first real applications of medical
visualization and remains important to this day. In 1993, Altobelli {\em et
al.}\ published their work on using CT data to visualize the possible outcome
of complicated craniofacial surgery~\cite{altobelli_computer-assisted_1993}.
By manually repositioning soft tissue fragments based on the bony surfaces
under them, in certain cases taking into account bone-skin motion ratios from
literature, the expected outcome of a craniofacial procedure could be
visualized. Although still rudimentary, this could be considered
one of the earliest cases of {\em predictive or outcome simulation}
integrated with visualization for surgical planning. The idea of predictive
simulation, or predictive medicine, was further explored by Taylor {\em et
al.}\ for cardiovascular surgery~\cite{taylor_predictive_1999}.

With the introduction of virtual colonoscopy (VC) in
1995~\cite{lichan_hong_3d_1995}, medical visualization also gained
diagnosis as an important medical application, namely screening for
colon cancer. VC combines CT scanning and volume visualization
technologies. The patient abdomen is imaged in a few seconds by a
multi-slice CT scanner. A 3D model of the colon is then reconstructed
from the scan by automatically segmenting the colon and employing
``electronic cleansing'' of the colon for computer-based removal of the
residual material. The physician then interactively navigates through
the volume rendered virtual colon employing camera control mechanisms,
customized tools for 3D measurements, ``virtual biopsy'' to interrogate
suspicious regions, and ``painting'' to support 100\% inspection of the
colon surface~\cite{hong_virtual_1997}.  VC is rapidly gaining
popularity and is poised to become the procedure of choice in lieu of
the conventional optical colonoscopy for mass screening for colon polyps
 -- the precursor of colorectal cancer. Unlike optical colonoscopy, VC
 is patient friendly, fast, non-invasive, more accurate, and cost-effective
procedure for mass screening for colorectal cancer. 

VC technologies gave rise to the computer-aided detection (CAD) of
polyps, where polyps are detected automatically by integrating volume
rendering, conformal colon flattening, clustering, and ``virtual
biopsy'' analysis. Along with the reviewing physician, CAD provides a
second pair of ``eyes'' for locating polyps~\cite{hong_pipeline_2006}.
This work was also the basis for many other virtual endoscopy systems,
such as virtual bronchoscopy, virtual cystoscopy, and virtual
angioscopy. A careful integration of image analysis (e.g., segmentation,
skeletonization), with efficient rendering (e.g., occlusion culling) and
interaction (e.g., camera control based on predefined paths) are major
ingredients of such systems~\cite{bartz_virtual_2005,hong_virtual_1997}.

\subsection{Multi-field Data}
\begin{figure}
\includegraphics[height=0.4\textheight]{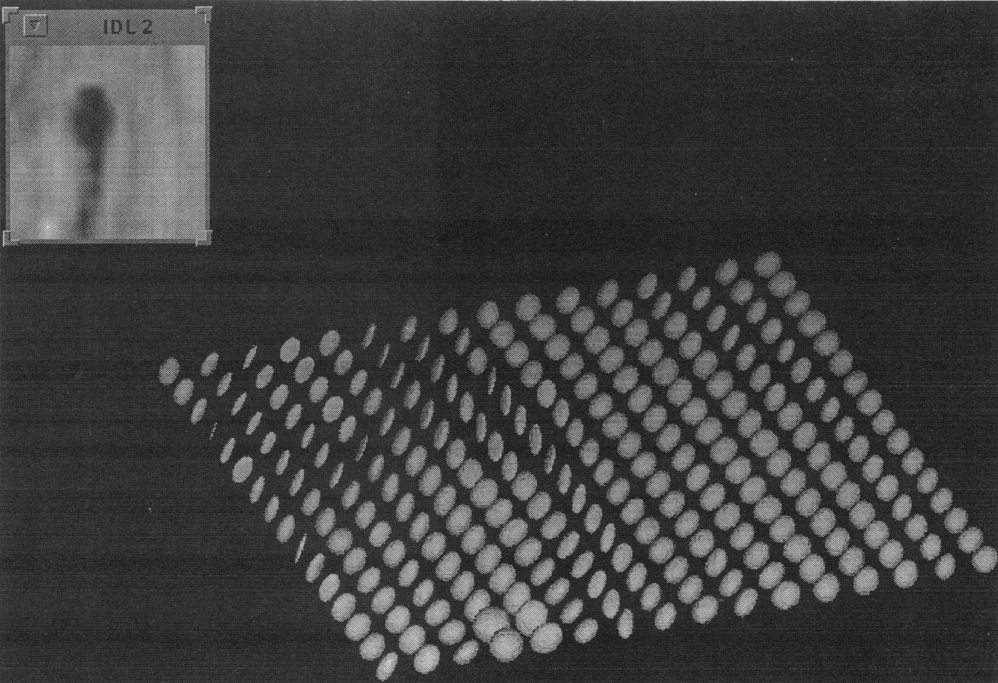}
\caption{Effective diffusion ellipsoid glyphs from a 2D region of
interest of a DTI dataset of a cat brain. Image taken
from~\cite{basser_mr_1994}.}
\label{fig:basser1994_dti_glyphs}
\end{figure}

Diffusion Tensor Imaging, or DTI, is an MRI-based acquisition modality,
introduced in 1994 by Basser {\em et al.}, that yields $3\times 3$ symmetric
diffusion tensors as its native measurement quantity~\cite{basser_mr_1994}. The
tensors represent the local diffusion of water molecules, and hence indirectly
indicate the presence and orientation of fibrous structures, such as
neural fiber bundles or muscle fibers. Already in this first paper, the authors employed 3D
glyphs to visualize the eigensystems represented by the tensors, as
shown in figure~\ref{fig:basser1994_dti_glyphs}. 

Basser and his colleagues were also some of the first to extract and visualize
fiber-tract trajectories from DTI data of the brain~\cite{basser_vivo_2000,
basser_fiber-tractography_1998}, thus linking together the point diffusion
measurements to get an impression of the global connective structures in the
brain. With DTI it was encouraging to see that the first visualization efforts
were initiated by the scientists developing this new scanning modality
themselves. Early work by the visualization community includes tensor lines for
tractography~\cite{weinstein_tensorlines:_1999} and direct volume rendering of
DTI data~\cite{kindlmann_hue-balls_1999, kindlmann_strategies_2000}.

Importantly, DTI serves as one of the first examples of natively multi-field
medical data, that is medical data with multiple parameters defined over the
same spatio-temporal domain. The advent of DTI initiated a whole body of
medical visualization research dedicated to the question of how best to
visually represent and interact with diffusion tensor data in particular and
multi-field medical data in general. The 2007 paper by Blaas {\em et al.}\
presented a visual analysis-inspired solution to this problem based on linked
physical and feature space views~\cite{blaas_interactive_2007}.

\subsection{Time-varying Data}
Time-varying medical volume data visualization made its entrance in 1996
with work by Behrens {\em et al.}\ on supporting the examination of
Dynamic Contrast-Enhanced MRI mammography data with the display of
parameter maps, the selection of regions of interest (ROIs), the
calculation of time-intensity curves (TICs), and the quantitative
analysis of these curves~\cite{behrens_computer-assisted_1996}. In 2001,
Tory {\em et al.}\ presented methods for visualizing multi-timepoint (1
month interval) MRI data of a multiple sclerosis (MS) patient, where the
goal was to study the evolution of brain white matter lesions over
time~\cite{tory_4d_2001}. Methods used included glyphs, multiple
isosurfaces, direct volume rendering and animation.
Coto {\em et al.}\ applied multiple coupled views, including linked
cursors and brushing on 3D renderings and scatterplots, to dynamic
contrast-enhanced MRI (DCE-MRI) mammography
data~\cite{coto_mammoexplorer:_2005}.

\subsection{Illustrative Visualization}
Illustrative visualization is primarily motivated by the attempt to create
renditions that consider \emph{perceptual capabilities} of humans. As an
example, humans infer information about shapes not only from realistic shading
but also from appropriate hatching and from outlines that support the mental
separation of nearby objects rendered in similar colours.

Illustrative visualization is related to the term
\textit{Non-Photorealistic Rendering} in computer graphics, or NPR for
short. The term NPR was used since around 1990 when the seminal paper of
Saito {\em et al.}\ clearly illustrated that complex 3D shapes could be
rendered more \emph{comprehensible} by using certain \emph{feature
lines}~\cite{saito_comprehensible_1990}. Compared to NPR, illustrative
visualization is the more focused term that covers rendering techniques
serving clear visualization goals, namely to convey shape information
efficiently.  In medical visualization, either surfaces or volume data
are rendered in illustrative styles. For illustrative volume rendering,
the term \emph{volume illustration} was introduced by Ebert {\em et al.}
in 2000~\cite{ebert_volume_2000}. Boundary enhancement based on gradient
approximation~\cite{csebfalvi_fast_2001} and curvature-based transfer
functions~\cite{kindlmann_curvature-based_2003} are landmarks in
illustrative medical visualization. Tietjen {\em et al.} applied
silhouettes and other feature lines for various scenarios in liver
surgery planning~\cite{tietjen_combining_2005}. Besides silhouettes,
stippling and probably even more hatching has great potential to reveal
details of shapes~\cite{interrante_enhancing_1995}.

Later, Bruckner {\em et al.}\ made a number of important contributions that
support depth and shape perception with adapted transfer functions. In
particular, they considered the peculiarities of interactive exploration of 3D
datasets and elaborated on the idea of preserving essential context
information~\cite{bruckner_illustrative_2005, bruckner_illustrative_2006,
bruckner_enhancing_2007}. These and later refinements are integrated in the
VolumeShop-system that is publicly available and used by several research
groups

\subsection{Multi-subject Data}
Medical visualization has also started to work on the problem of dealing with
multi-subject data. These are datasets that include measurements,
including imaging, of more than one subject. The goal is to be able to extract
patterns that affect sub-groups of the whole collection, for example to explore
which aspects of the data correlate with a specific disease outcome. Examples of
this type of work include LifeLines2, an information visualization
approach to visualize and compare multiple patient histories or electronic medical
records~\cite{wang_aligning_2008}. More recently, work has been done on the
interactive visualization of the multi-subject and mixed modality datasets
acquired by medical cohort studies~\cite{steenwijk_integrated_2010}. In these
studies, mixed modality data, including imaging, genetics, blood measurements
and so on, is acquired from a group of subjects in order to understand, diagnose
or predict the clinical outcome of that group. Steenwijk et al.\ demonstrated
that it was possible to create a highly interactive coupled view visualization
interface, integrating both information and scientific visualization techniques,
with which patterns, and also hypotheses, could be extracted from the whole data
collection.

%



\section{Challenges in Medical Visualization}

\subsection{Advances in Data Acquisition}
Toshiba's 320-slice CT scanner, the Aquilion One, was introduced in 2007. It is
able to acquire five 320 slice volumes {\em per second}~\cite{hsiao_ct_2010} and
can thus image a beating heart. Rapid and continuous advances in the dynamic
nature and sheer magnitude of data in this and other mainstream medical imaging
necessitates improvements to existing techniques in terms of computational and
perceptual scalability.

High Angular Resolution Diffusion Imaging (HARDI)~\cite{tuch_high_2002} and
Diffusion Spectrum Imaging (DSI)~\cite{hagmann_understanding_2006} datasets
contain hundreds of diffusion-weighted volumes describing the diffusion of water
molecules and hence indirectly the orientation of directed structures such as
neural fiber bundles or muscle fibers. This is a rather extreme example of
multi-field medical data that is becoming more relevant in both medical research
and clinical application. Completely new visual metaphors are required to cope
with the highly multi-variate and three-dimensional data of diffusion weighted
imaging in particular and many other new imaging modalities in general.

Molecular imaging enables the {\em in vivo} imaging of biochemical
processes at macroscopic level, meaning that, for example, pathological
processes can be studied and followed over time in the same subject long
before large-scale anatomical changes occur. Examples are
bioluminescence (BLI) and fluorescence (FLI) imaging, two molecular
imaging modalities that enable the {\em in vivo} imaging of gene
expression.  Molecular imaging yields datasets that vary greatly in
scale, sensitivity, spatial-temporal embedding and in the phenomena that
can be imaged. Each permutation brings with it new domain-specific
questions and visualization challenges. Up to now, most of the
visualization research has been focused on small animal
imaging~\cite{kok_integrated_2007, kok_articulated_2010}, but due to its
great diagnostic potential, molecular imaging will see increased
application in humans.

The integration of microscopy imaging is an essential task for the future,
where data handling, interaction facilities but also more classical rendering
tasks such as transfer function design become essential. With more and more
large scale and 3D microscopy data available, there are many opportunities for
visualization researchers. Recent examples include techniques for
interactively visualizing large-scale biomedical image stacks demonstrated on
datasets of up to 160 gigapixels~\cite{won-ki_jeong_interactive_2010} and tools
for the interactive segmentation and visualization of large-scale
3D neuroscience datasets, demonstrated on a 43 gigabyte electron microscopy
volume dataset of the hippocampus~\cite{jeong_ssecrett_2011}.

With these examples, we hope to have illustrated that advances in image
acquisition are continuous, and due to the increasing demands of modern society
are accelerating. Each new advance in imaging brings potentially greater
magnitudes and exotic new types of data leading to new challenges for medical
visualization.

\subsection{Heterogeneous Display and Computing Devices}
Mobile devices, in particular the \textsc{Apple} products \textsc{iPad} and
\textsc{iPhone}, are extremely popular among medical doctors and indeed
solve some serious problems of desktop devices in routine clinical use. In
particular, bedside use of patient data is an essential use case for medical
doctors of various disciplines.

Meanwhile several mobile devices are equipped with powerful graphics cards and
using the OpenGL ES (Embedded Systems) standard, they are able to provide
high-quality interactive rendering. Although the performance still tails that
of modern desktop devices, slicing medical volume data and 3D rendering is
feasible~\cite{moser_interactive_2008}.

The rapid and widespread use of mobile devices also made gesture input popular.
In particular, multi-touch interaction is considered an intuitive interaction
since many potential users know a variety of gestures from their everyday
activities with smart phones. Therefore multitouch interaction is also
incorporated in large displays in medical use, e.g. the Digital
Lightbox\footnote{\url{http://www.brainlab.com/art/2841/4/surgical-pacs-access/}}
by BrainLab and the multi-touch table of Lundstr\"om {et
al.}~\cite{lundstrom_multi-touch_2011}.

\subsection{Interactive Image Segmentation}
Image segmentation is important in clinical practice, for example,
diagnosis and therapy planning, and also in image-based medical research. In these
applications, segmentation is complicated by the great deal of variation in
image acquisition, pathology and anatomy. Furthermore, in matters of diagnosis
or therapy planning, the accuracy of the segmentation can be critical. It comes
as no surprise that user interaction is often required to help ensure the
quality of the results, by initializing an image processing method, checking the
accuracy of the result or to correct a
segmentation~\cite{olabarriaga_interaction_2001}.

A well-known interactive segmentation technique is the live-wire or intelligent
scissors~\cite{mortensen_intelligent_1995}. These ideas were later extended and
applied to medical images~\cite{falcao_user-steered_1998}. More recently,
visualization has been applied to the challenge of explicitly dealing with the
uncertainty inherent in interactive 3D
segmentation~\cite{prassni_uncertainty-aware_2010, saad_exploration_2010}.

Medical visualization research often combines elements of image analysis,
graphics and interaction, and is thus ideally equipped to address the challenge
of developing and validating effective interactive segmentation approaches for
widespread use in medical research and practice.

\subsection{Topological Methods}

Topological data representation has played an important role in medical
visualization, since it can allow us to segment specific features such as human
organs and bones from the input medical datasets systematically and identify
their spatial relationships consistently. Actually, such topological concepts
have been already introduced in the earlier stage of medical visualization. For
example, the 3D surface of a human cochlea was reconstructed from a
series of 2D CT cross-sectional images by identifying correct correspondence between the
cross-sectional contours~\cite{shinagawa_constructing_reeb-1991}.

\begin{figure}
\centering{\includegraphics[height=0.3\textheight]{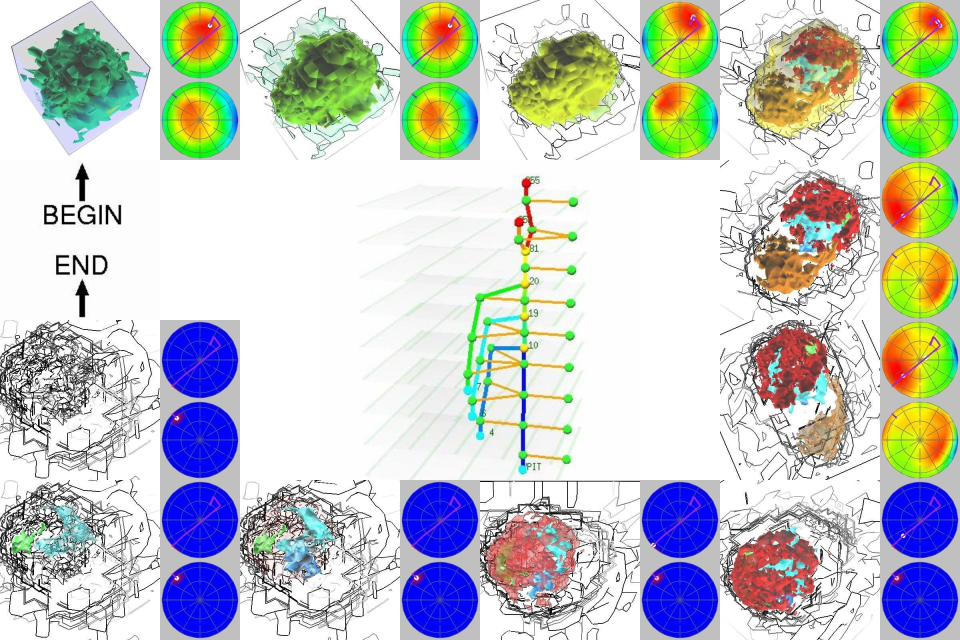}}
\caption{Previewing a sheep heart volume along the optimal camera path. Image taken from~\cite{takahashi_previewing_volume_2011}.}
\label{fig:previewing_volume}
\end{figure}

Topological approaches have also been extended to analyze 3D medical
volume data. Contour trees~\cite{bajaj_contour_spectrum_1997} have been
employed for designing transfer functions in order to illuminate human
organs and bones systematically since the associated anatomical
structure can be effectively captured as topological skeletons of
isosurfaces~\cite{weber_topology_controlled_2007}. Spatial relationships
between bones and the position of aneurysm were successfully extracted
respectively from CT and angiographic datasets using a variant of
contour tree~\cite{chan_relation_aware_2008}. Interesting features in
medical volume data can be visually analyzed using an optimal camera
path, which can be obtained by referring to the topological structure of
human organs~\cite{takahashi_previewing_volume_2011} (see
figure~\ref{fig:previewing_volume}). Topological methods are now being
developed for visualizing multi-variate and high-dimensional datasets,
and thus potentially for analyzing tensor fields obtained through
DT-MRI, multi-subject data in group fMRI studies, and time-varying data measured
by high-speed CTs.

\subsection{Integration of Simulation Models}
In their 1999 predictive medicine paper, Taylor {\em et al.}\ argued
that surgical planning should not only address questions of surgical
approach but also of the expected outcome, for example predicted future
states such as the efficacy of a treatment option or the performance of
an implant~\cite{taylor_predictive_1999}. Medical visualization
approaches become significantly more valuable when enhanced with
simulation models that help to predict the outcome of a disease process
or therapeutic procedure, or that enrich measured data with expected
physiological phenomena. Examples besides the blood flow simulations of
Taylor {\em et al.} include interactive skeletal range of
motion~\cite{krekel_interactive_2006} and biomechanical
stress~\cite{dick_stress_2009} simulation models for implant planning in
orthopedics and nasal airflow simulation for reconstructive
rhinosurgery~\cite{zachow_visual_2009}.

The integration of these predictive models, although potentially valuable,
brings with it new challenges. The addition of potentially complex and dynamic
simulation output data to existing visualizations requires new visual
representation techniques. Furthermore, for the simulation results to be
maximally useful, the models should be tightly coupled to and steered by the
user's interaction with the medical visualization. Finally, most simulations
yield data with a certain degree of inherent uncertainty. The role of this
uncertainty should be fully explored and it should be carefully but explicitly
represented as an integral part of the visualization.

\subsection{Mappings and Reformations}
\begin{figure}
\includegraphics[height=0.28\textheight]{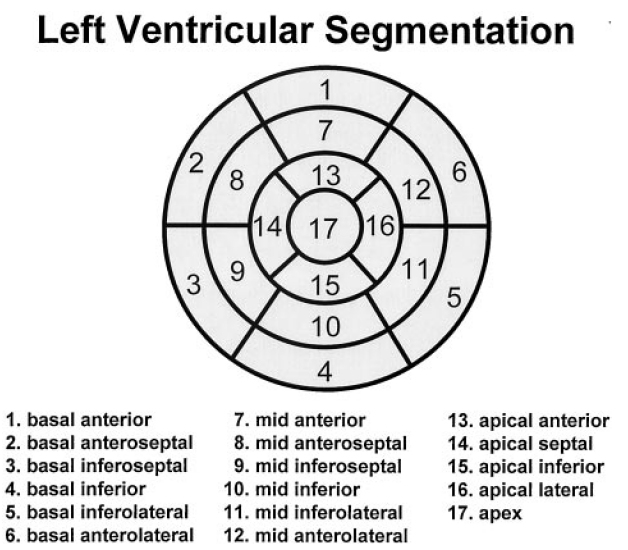}
\includegraphics[height=0.28\textheight]{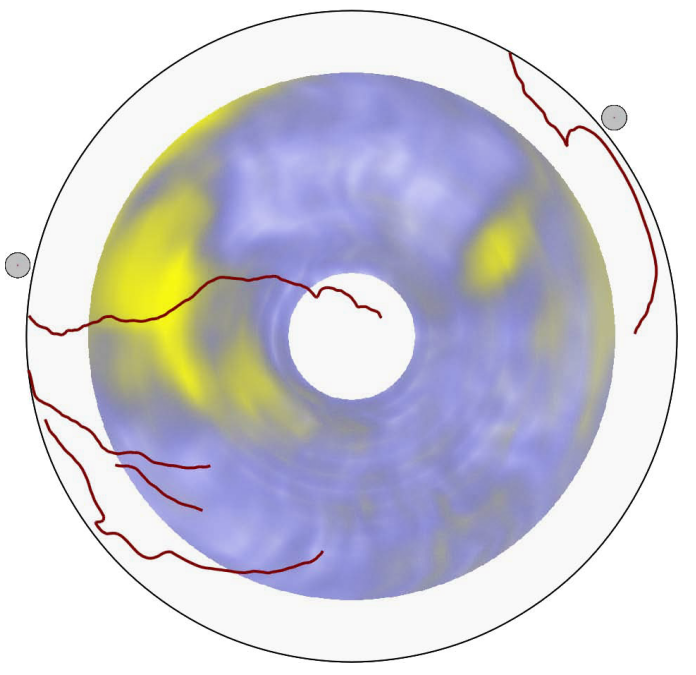}
\caption{On the left, the American Heart Association standardized 2D
bull's eye plot (BEP) of the left ventricle of the
heart~\cite{cerqueira_standardized_2002}. On the right, the volumetric
bull's eye plot, a modernized version by Termeer et al.~\cite{termeer_covicad:_2007}.}
\label{fig:aha_bep_2002}
\end{figure}

In 2002, the American Heart Association proposed a standardised segmentation and
accompanying 2D bull's eye plot (see figure~\ref{fig:aha_bep_2002}) of the
myocardium, or heart muscle, of the left heart
ventricle~\cite{cerqueira_standardized_2002}. This 2D plot is a simple but great
example of reducing complex 3D data to a standardized 2D representation that
greatly facilitates the interpretation of that data. Another well-known example
is that of curved planar reformation, or CPR, where volume data is sampled along
a curved plane following the trajectory of a blood vessel or other tubular
structure, thus enabling the study of the vessel and its surroundings with the
minimum of interaction~\cite{kanitsar_cpr:_2002}. Other good examples of
reformation can also be found in brain flattening~\cite{fischl_cortical_1999}
and colon unfolding~\cite{vilanova_nonlinear_2001}.

Recently, the idea of intelligently reformatting or mapping 3D data was
further explored by Neugebauer {\em et al.}\ with aneurysm maps for the
visualization of complex blood flow simulation data on the inside
surfaces of aneurysms~\cite{neugebauer_map_2009} and by Rieder {\em et
al.}\ with their tumor maps for the post-operative assessment of
radiofrequency ablation therapy~\cite{rieder_visual_2010}. These types
of reformations and mappings entail that more effort has to be put into
carefully designing simplified, usually 2D, representations of complex
3D data, as opposed to for example the relatively straight-forward
projection of volume data. The resultant visualization, if done right,
requires less or no interaction and by definition avoids a number of
problems inherent in 3D representations~\cite{ware_designing_2001}.

\subsection{Illustrative Visualization in Medicine}
For a long time, it was not possible to apply illustrative visualization
techniques in practice due to performance constraints. With advances in graphics
hardware and algorithms, such as GPU
raycasting~\cite{kruger_acceleration_2003}, it is now feasible from a
computational standpoint. Now that computational problems have been
largely solved, illustrative visualization approaches have to be finetuned and evaluated for diagnostic and
treatment planning purposes. Recent examples of such work include the simulation
of crepuscular rays for tumor accessibility
planning~\cite{khlebnikov_crepuscular_2011} and multi-modal illustrative
volume rendering for neurosurgical tumor treatment~\cite{rieder_interactive_2008}.

Illustrative medical visualization becomes increasingly important when
visualizations become more complex and multi-modal, integrating functional
(measured and simulated), anatomical information and for example surgical
instruments. Illustration techniques enable visual representations to be
simplified intelligently by the visualization designer, whilst still
communicating as much information as possible. An example of this is the work of
Zachow {\em et al.}\ on the visualization of nasal air flow simulation where
boundary enhancement was used as an illustrative technique to convey the
simulated flow and the anatomy simultaneously~\cite{zachow_visual_2009}.

\subsection{Hyper-realism}

\begin{figure}
\includegraphics[height=0.37\textheight]{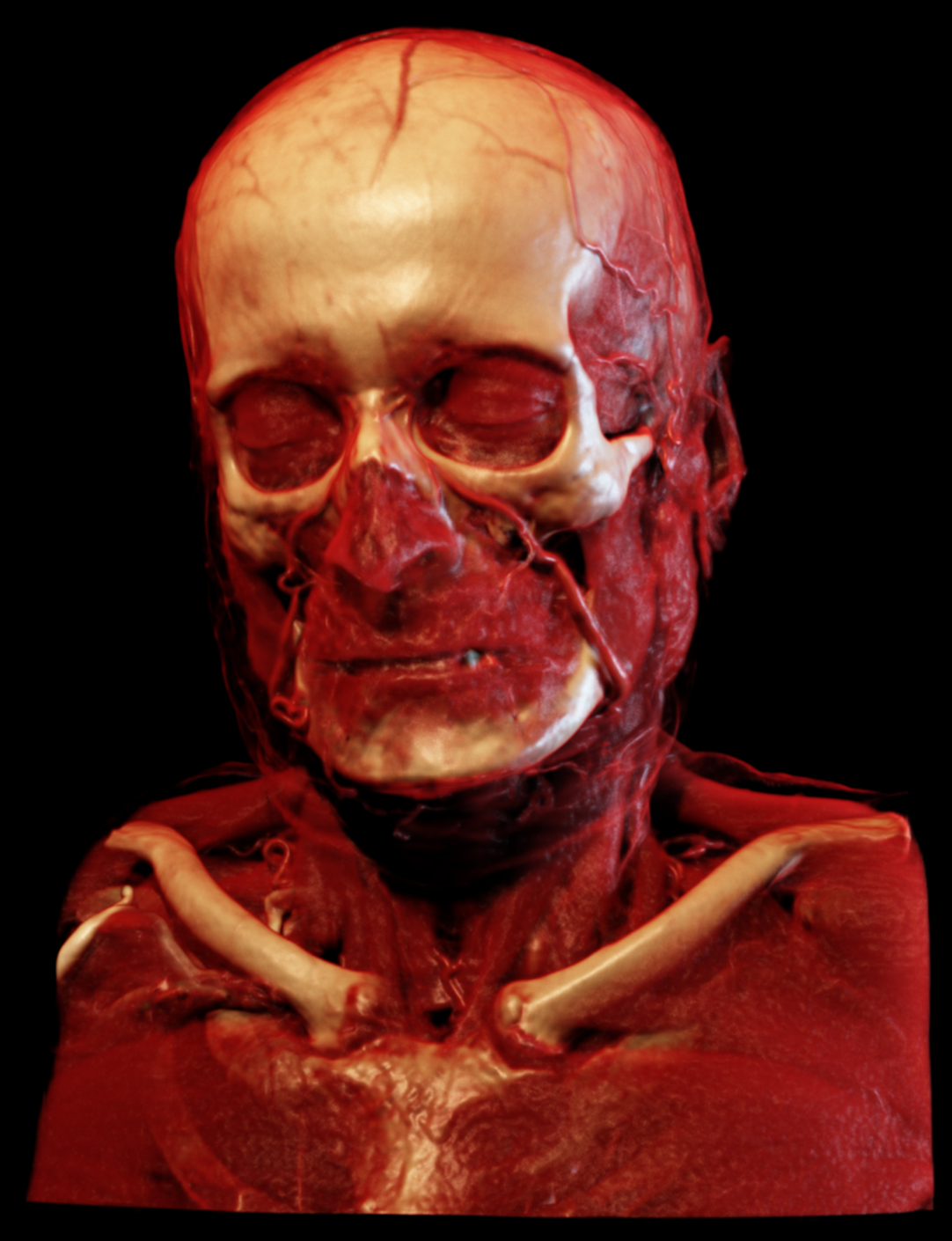}
\includegraphics[height=0.37\textheight]{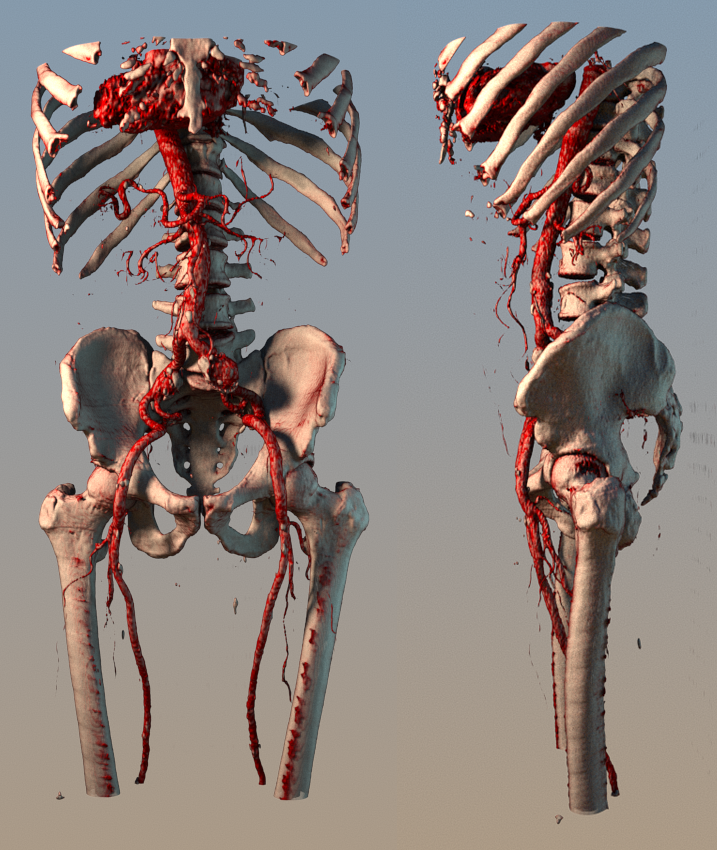}
\caption{Two examples of {\em interactive} visualizations made with the volume
renderer of Kroes {\em et al.}~\cite{kroes_exposure_2012}. Through the use
of GPUs, physically-based lighting has become possible in an interactive
volume rendering setting, enabling increased realism through soft shadows,
depth of field and in this case mixed phase function and BRDF surface
scattering.}
\label{fig:kroes2011}
\end{figure}

Analogous to the case of illustrative visualization, the rapid development
in graphics hardware and algorithms has now enabled the {\em interactive}
rendering of medical imaging datasets with physically-based
lighting~\cite{kroes_exposure_2012}. Figure~\ref{fig:kroes2011} shows an
example of such a visualization. These techniques make possible the simulation
of an arbitrary number of arbitrarily shaped and textured lights, real shadows,
a realistic camera model with lens and aperture, and so forth, all at
interactive rates.

These techniques enable not only photo-realism, but also a technical form of
{\em hyper-realism} in art, where it is possible to enhance visualizations with
{\em additional} realistic detail in order to better convey information. Whilst
there are strong indications that for example global illumination and shadows can have
a positive effect on task performance in normal volume
rendering~\cite{lindemann_about_2011}, the possibilities and value of
hyper-realistic effects in medical visualization need to be explored.

\subsection{Visual Analysis in Healthcare}
Visual analysis is becoming an essential component of medical visualization
due to the rapidly growing role and availability of complex multi-dimensional,
time-varying, mixed-modality, simulation and multi-subject datasets. In our
view, the magnitude and especially the heterogeneity of the data necessitate the
use of visual analysis techniques.

Existing examples involving time-varying data include the work of Coto {\em
et al.}\ on DCE-MRI mammography~\cite{coto_mammoexplorer:_2005} and
Oeltze {\em et al.}\ on perfusion data in general and MR perfusion of the brain in
particular~\cite{oeltze_interactive_2007}. Blaas {\em et al.} applied visual
analysis techniques to multi-modal medical data, whilst Zachow {\em et al.}\
focused on nasal airflow simulation data combined with anatomical
information~\cite{zachow_visual_2009}.

There is great potential for visual analysis in medical visualization, with
clinical applications including advanced diagnosis and medical research and,
even more importantly, treatment planning and evaluation, e.g. radio therapy
planning and post-chemotherapy evaluation. The new Visual Analysis in Healthcare
(VAHC) workshops that were held at IEEE VisWeek in 2010 and 2011 underline the
emerging importance of this research direction.

\subsection{Population Imaging}
In population imaging, medical image data and other measurements are acquired of
a large group of subjects, typically more than one thousand, over a longer
period, typically years, in order to study the onset and progression of disease,
general aging effects, and so forth in larger groups of people. Examples
include the Rotterdam Scan Study focusing on
neuro-degeneration~\cite{de_leeuw_prevalence_2001} and the Study of
Health In Pomerania (SHIP) focusing on general health~\cite{john_study_2001}.

This application domain is an extreme example of multi-subject medical
visualization discussed in section~\ref{sec:overview}, integrating
large quantities of heterogeneous, multi-modal and multi-timepoint data acquired
of a large group of subjects. The scientists running these studies usually do
not formulate strictly-defined hypotheses beforehand, instead opting for
meticulous data acquisition, followed by an extended period of analysis in order
to extract patterns and hypotheses from the data. Recently, Steenwijk {\em et
al.}\ set the first steps for the visualization of population imaging by
applying visual analysis techniques to cohort study imaging
data~\cite{steenwijk_integrated_2010}. The extreme heterogeneity and
magnitude of the data, coupled with the explorative nature of the research, renders this a promising long-term
application domain for visual analysis and medical visualization.

\section{Conclusions}

In this chapter, we gave a compact overview of the history of medical
visualization research, spanning the past 30 years. Based on this history and
on our own observations working in the field, we then identified and discussed
the research challenges of the coming decade.

Our discussion of classic medical visualization problems related to efficient
and high quality display of \emph{one} static dataset was brief. We devoted
more space to data that change over time, to the integration of anatomy with
simulation and finally to cohort studies. We refer to problems where such
time-dependent and high-dimensional data are employed as \emph{MedVis 2.0}
problems. While the classic problems are -- from an application perspective --
solved, there are many research opportunities in MedVis 2.0 problems. These
data are significantly more difficult to analyze, to process and to visualize.
Time-dependent MRI data, e.g., exhibit all artifacts of static MRI data but a
number of additional artifacts, e.g. due to motion. Integrated analysis and
visualization is a key feature of MedVis 2.0 solutions. In general, successful
solutions to these problems require a considerably deeper understanding of the
medical background and thus favor close collaborations with medical doctors
over merely having access to medical data.

\bibliographystyle{spmpsci}

\bibliography{medvis_challenges}

\end{document}